\documentclass[floatfix,showpacs,pre,aps,a4paper,twocolumn]{revtex4}

 \usepackage{bm}
 \usepackage{amsmath}
 \usepackage{amssymb}
 \usepackage{dsfont}
 \usepackage{latexsym}
 \usepackage{amsfonts}
 \usepackage{epsfig}
 \usepackage{color}
 \definecolor{darkblue}{rgb}{0,0,.5}
 \usepackage[linktocpage, colorlinks=true, linkcolor=darkblue, citecolor=darkblue]{hyperref}
 \usepackage[all]{hypcap}

\newcommand{\C}[1]{{\cal{#1}}}
\newcommand{\bb}[1]{\textbf{#1}}
\newcommand{\lr}[1]{{\langle {#1}\rangle}}

\begin{document}

\title{Thermodynamics of stochastic Turing machines}

\author{Philipp Strasberg, Javier Cerrillo, Gernot Schaller, and Tobias Brandes}
\affiliation{Institut f\"ur Theoretische Physik, Technische Universit\"at Berlin, Hardenbergstr. 36, D-10623 Berlin, Germany}

\begin{abstract}
 In analogy to Brownian computers we explicitly show how to construct stochastic models, which mimic the behaviour of 
 a general purpose computer (a Turing machine). Our models are discrete state systems obeying a Markovian master 
 equation, which are logically reversible and have a well-defined and consistent thermodynamic interpretation. The 
 resulting master equation, which describes a simple one-step process on an enormously large state space, allows us 
 to thoroughly investigate the thermodynamics of computation for this situation. Especially, in the stationary regime 
 we can well approximate the master equation by a simple Fokker-Planck equation in one dimension. We then show that the 
 entropy production rate at steady state can be made arbitrarily small, but the total (integrated) entropy 
 production is finite and grows logarithmically with the number of computational steps. 
\end{abstract}


\maketitle

\section{Introduction}

Computers are physical systems and information has to be stored and transmitted using physical devices. 
This trivially sounding statement has led to very important insights as soon as one starts to ask 
for the fundamental physical limits of computation. The most known statement is probably Landauer's principle: 
erasing a data set with information content (i.e., entropy) $H$ causes a minimum heat dissipation of 
$\beta^{-1} H$ if $\beta$ is the inverse 
temperature of the surrounding environment. This was first formulated by Landauer in 1961~\cite{Landauer1961}. 
More generally, it was argued by Bennett~\cite{BennettIBM1973, Bennett1982, BennettIBM1988} and others 
\cite{KeyesLandauerIBM1970, Likharev1982, BennettLandauerSciAm1985, Feynman1985, LeffRexBook}, that each logically 
irreversible operation (as, e.g., information erasure) must be accompanied with a corresponding heat dissipation 
whereas \emph{each logically reversible operation can be implemented -- at least in principle -- in an energetically 
neutral way.} The heat flow of such computers, if operated slowly enough, is then equal to the change in their Shannon 
entropy (times $\beta^{-1}$) demonstrating the usefulness of the Shannon entropy to describe thermodynamic processes. 

Although the question about the relation between logical and thermodynamical reversibility sounds like a purely 
academic question, it is indeed of practical importance because one of the limitations of todays computers lies 
in the heat which they produce during computation and which is hard to be drained off quickly enough. 
In addition, it is well-known that the thermodynamics of computation can be successfully applied to 
resolve the famous Maxwell demon paradox~\cite{Bennett1982, BennettIBM1988, LeffRexBook}. 

Today, it seems that most physicists accept the thermodynamics of computation as a well established field and 
also Feynman notes (page 160 in~\cite{Feynman1985}): ``I see nothing wrong with his [Bennett's] arguments. [...] I 
concluded that there was no minimum energy [consumption of computers].'' 
Indeed, however, criticism was raised against Bennett's exorcism of Maxwell's demon and the thermodynamics of computation 
already some time ago~\cite{EarmanNortonStudHistPhil1998, EarmanNortonStudHistPhil1999, ShenkerStudHistPhil1999}, which 
was subsequently defended by Bub and Bennett again~\cite{BubStudHistPhil2001, BennettStudHistPhil2003}, 
but criticism and controversies still prevail for different reasons, mainly (but not only) on the philosophical side 
\cite{MaroneyStudHistPhil2005, NortonStudHistPhil2005, LadymanEtAlStudHistPhil2007, HemmoShenkerJPhilos2010, NortonStudHistPhil2011, KishGranqvistEPL2012, NortonFoundPhys2013, NortonEntropy2013, KishEtAlConfSer2014, AlickiArXiv1, AlickiArXiv2, NortonIntJModPhys2014}. 

An important class of physical models used to illustrate the thermodynamics of computation are inspired by biochemical 
processes as, e.g., DNA replication~\cite{BennettIBM1973, Bennett1982, BennettLandauerSciAm1985, Feynman1985}. Indeed, 
copying a DNA strand can be regarded as a simple computational task where the DNA strand represents a certain input 
signal, which is manipulated by enzymes to produce an identical copy of the input. Energetic barriers in the 
computational path, i.e., barriers between two logical states of the computation, can be overcome by the random thermal 
motion of the molecules involved and a bias in chemical potentials can be used to drive the computation in a desired 
direction. For a small enough chemical bias the average dissipation of energy per step can be made arbitrarily small and 
thus, one usually concludes that computation can be carried out thermodynamically reversibly. 

In a more general frame, systems which use the random thermal motion of its components to perform a computation are 
usually called \emph{Brownian computers}. However, in addition to the arguments presented above, a detailed and general 
mathematical treatment of such computers seems to be missing in the literature. Indeed, the authors of Refs. 
\cite{BennettIBM1973, Bennett1982, BennettLandauerSciAm1985, Feynman1985} based their reasoning largely on ingenious 
arguments instead of detailed calculations. If one finds more detailed (yet not very general calculations) in the 
literature~\cite{NortonFoundPhys2013, NortonIntJModPhys2014}, then they seem to contradict the statement that Brownian 
computers are thermodynamically reversible. 

In the present contribution we will therefore treat the subject of Brownian computers in general, i.e., without having any 
specific computational task or physical problem in mind. We will start by considering an arbitrary Turing machine (TM), 
which is known to be a model for a general purpose computer. This means that for each computable function (or algorithm) 
there exists a TM which can implement it. Moreover, it is even possible to construct a special TM, called 
a universal TM, which is able to simulate any other TM and hence, TMs are said to be computationally universal (an 
introduction to TMs can be found in Refs.~\cite{Feynman1985, MinskyBook}). Based on the ideas of Bennett we will then show 
how to construct a logically reversible TM and in addition, we show how to model this TM by a continuous-time Markov 
process, 
i.e., by a Markovian master equation (ME). The resulting model is then able to compute in a stationary regime where it 
transforms a string of incoming symbols (the inputs of the computation) into a corresponding string of outgoing symbols 
(simply called the outputs). 

The big advantage of a description in terms of a ME is that its thermodynamic behaviour is well understood since many 
years~\cite{SchnakenbergRMP1976, VanKampenBook}, and also their stochastic behaviour can be treated within a consistent 
thermodynamic formalism, which is known as stochastic thermodynamics~\cite{SeifertRepProgPhys2012}. 
Indeed, there has been a large interest recently in using small autonomous machines describable by, e.g., a Markovian ME, 
to address questions of information processing as, e.g., sensing, feedback or adaptation, in a thermodynamic context. 
Such machines were studied for abstract 
models~\cite{AllahverdyanEtAlJStatMech2009, MandalJarzynskiPNAS2012, BaratoSeifertEPL2013, MandalQuanJarzynskiPRL2013, MunakataRosinbergJStatMech2013, ShiraishiEtAlNJP2015}, 
more general 
settings~\cite{DeffnerJarzynskiPRX2013, HorowitzEspositoPRX2014, HartichJStatMech2014, BaratoSeifertPRE2014, ShiraishiSagawaPRE2015}, 
in a biochemical 
context~\cite{AllahverdyanWangPRE2013, BaratoPRE2013, BaratoNJP2014, SartoriEtAlPLoS2014, SartoriPigolotti} or 
in the field of artificial 
nanostructures~\cite{StrasbergEtAlPRL2013, StrasbergEtAlPRE2014}. 
Furthermore, we have reached the realm where experiments are performed at the Landauer 
limit~\cite{BerutNature2012, JunPRL113, HongEtAlArXiv2014, SilvaEtAlArXiv2014, BerutArXiv2015}. 
However, to the best of our knowledge, there is no reference which has treated the question of how to describe a 
\emph{general} computer within a ME framework and which has worked out its thermodynamic consequences in detail. 
We will find out that the entropy production rate of a Brownian computer can be made arbitrarily small while the total 
entropy production grows logarithmically with the number of computational steps, thereby resolving a part of the 
controversy between Bennett, Norton and others. 

\emph{Outline: } Because the treatment of TMs is no standard subject taught in physics, we will give an introduction 
to it in Sec.~\ref{sec TMs} to make the paper as self-contained as possible. Then, in Sec.~\ref{sec stochastic TM} 
we will first show how to build a 
TM in a logically reversible way before we demonstrate how to map it to a ME. Because the details of a logical reversible 
TM are a bit technical but not of major importance for a general understanding of the rest of the paper, we will shift 
them to appendix~\ref{app logical reversible TM}. Finally, after we have discussed the structure of the ME 
in Sec.~\ref{sec stochastic TM}, we discuss the thermodynamics of Brownian computation in Sec. 
\ref{sec thermodynamics Brownian computation}. A last section is then devoted to a summary of our 
results and an outlook on interesting future work.

\section{Turing machines}
\label{sec TMs}

\begin{figure}
 \includegraphics[width=0.25\textwidth,clip=true]{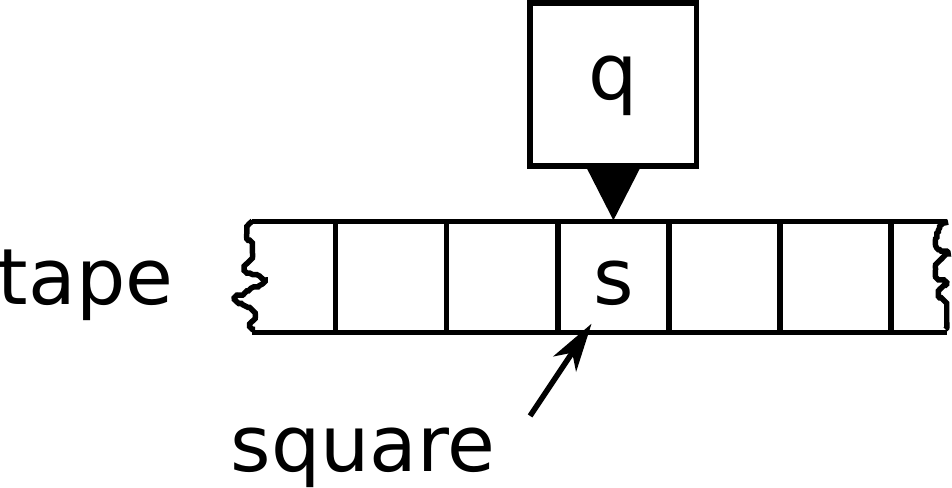}
 \caption{\label{fig TM sketch} Simple sketch of a TM: the machine, specified by a state $q$, has access to an infinite tape, which is divided 
 into equal squares. The machine scans with its head always only one square on the tape, on which it finds a symbol 
 $s$ (which might be also the blank symbol $b$) written. }
\end{figure}

A TM $T$ is an idealized machine to model or simulate problems in computer science (see Fig.~\ref{fig TM sketch}). 
In the standard treatment $T$ 
has two parts: first, we have the machine itself, which can be in some state $q\in\C Q$ where $\C Q$ denotes the finite 
set of internal machine states. Second, the machine has access to an external storage medium (usually called the 
\emph{tape}), which is divided into equal \emph{squares} and each square contains either a special symbol $s\in\C S$ (where 
$\C S$ is again a finite set) or it contains a blank $b$. The machine has a \emph{head} with which it is coupled to one 
and only one square of the tape at each point of time. 

At each time step, the machine in state $q$ reads the symbol $s$ written on the square and subsequently it changes its own 
state to $q'$, writes a new symbol $s'$ on the square (which can be the same as the old one) and either shifts the tape 
one square to the left or to the right or stays where it is. Mathematically, these rules can be defined by three functions 
$F, G$ and $H$: 
\begin{equation}\label{eq TM standard map}
 \begin{split}
  q'	&=	F(q,s),	\\
  s'	&=	G(q,s),	\\
  d'	&=	H(q,s)
 \end{split}
\end{equation}
where $d'\in\{-1,0,+1\}$ encodes the direction of movement of the tape with $-1,0,+1$ meaning move the tape right, do not 
move it, move it left. These three maps (together with an agreement about the initial state, see below) 
\emph{completely define} the action of $T$. Often one writes these maps in terms of a so-called quintuple 
\begin{equation}
 (q,s,F(q,s),G(q,s),H(q,s)) = (q,s,q',s',d')
\end{equation}
or simply as 
\begin{equation}
 (q,s) \rightarrow (q',s',d').
\end{equation}

A computation of $T$ is then defined as follows: the machine starts initially in a special state R $\in\C Q$ 
(we use R for ``ready'') and scans by convention the first blank symbol to the left of a finite number of input 
symbols $\bb s_\text{in} \equiv (s_1,s_2,\dots,s_M)$ written on the tape. Note that we demand that there is no blank symbol in 
between the input symbols and that the input string is finite. Then, $T$ starts to move right and reads the first 
input symbol $s_1$. It then proceeds according to the rules above [Eq. (\ref{eq TM standard map})]. 
After some time the TM might be done with the computation. We then assume that it shifts to the first blank symbol to the 
right of the remaining string of symbols $\bb s_\text{out} \equiv (s_1,s_2,\dots, s_{M'})$ written on the tape and then 
changes to a special final state H (H for ``halt'').
The string $\bb s_\text{out}$ is called the output or 
the \emph{result} of the computation, which is again finite but not necessarily of the same length as the input 
(i.e., $M'\neq M$ is possible). In short, we will also write a computation as 
$T: \bb s_\text{in} \rightarrow \bb s_\text{out}$ or $T(\bb s_\text{in}) = \bb s_\text{out}$. 

Hence, we see that the idea of a TM is very simple: given $(T,\bb s)$, i.e., a TM $T$ with input $\bb s$ in standard 
format as above, it follows the rules (\ref{eq TM standard map}) until it is done 
with the computation. Especially, we note that there is only a \emph{finite} set of quintuples or rules 
(\ref{eq TM standard map}) because $\C Q$ and $\C S$ were assumed to be finite sets. 
Introducing $N_{\C Q} \equiv \#\C Q$ 
and $N_{\C S} \equiv \#\C S$ (with ``$\#$'' denoting the cardinality of a set), we see that any TM is completely specified 
by $N_{\C Q}(N_{\C S}+1)$ many quintuples. Note that we also need a rule for the machine if it scans a blank symbol, 
hence the factor $N_{\C S}+1 = \#(\C S \cup\{b\})$. 

In view of these facts it seems very remarkable that, first of all, TMs are capable of universal computation and, 
second, that they can show an incredibly complex behaviour. The first property is related to the Church-Turing thesis 
-- which has to be taken for granted though -- which states that every intuitvely computable function can be computed by a 
TM. The second property is reflected, for instance, in the fact that it is \emph{impossible} to design a TM $T_H$, which 
tells us for an arbitrary given TM $T$ and input $\bb s$ whether $(T,\bb s)$ will halt or not. This is the famous halting 
problem. Thus, it might be that the computation defined above never reaches the state H and goes on forever. In this case, 
the computation simply has no result. Thus, the incredible power of TMs (namely their computational universality) has 
a serious drawback (namely their in general unpredictable behaviour). A much more detailed account of TMs 
can be found in Refs.~\cite{Feynman1985, MinskyBook}.

\section{Stochastic Turing machines}
\label{sec stochastic TM}

\subsection{Setup and idea}

\begin{figure}
 \includegraphics[width=0.45\textwidth,clip=true]{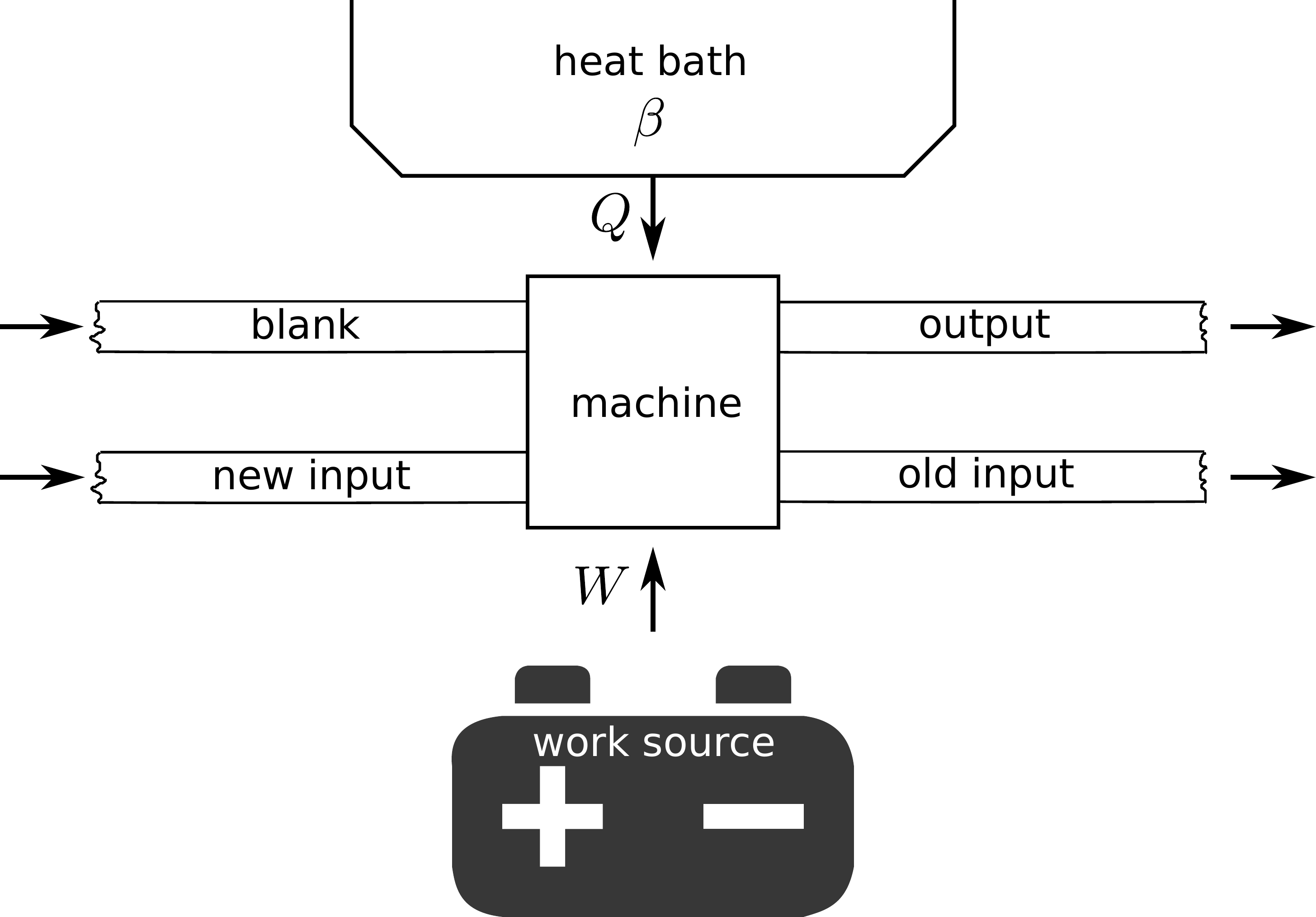}
 \caption{\label{fig TD comp} Sketch of the general setup, which allows us to analyze any abstract computational process in terms of 
 thermodynamic quantities. The upper tape corresponds to the output tape, which is initially blank, and the lower tape 
 corresponds to the input tape. Note that we will always assume that blank tapes are for free, i.e., the machine can have 
 as many blank space on which it can write as it requires. Furthermore, the machine itself has access to two additional 
 internal tapes (not sketched), see Sec.~\ref{sec logical reversible TM}. }
\end{figure}

Examining the thermodynamics of computation can be done in many different ways. Here, as stressed in the introduction, 
we want to capture three main features: first, we want to look at a general computational problem and not one specific 
task; second, we are interested in a logically reversible computer; and third, we want to model the computation 
stochastically, i.e., as a Brownian computer. The general thermodynamic picture is hence as sketched in Fig. 
\ref{fig TD comp}. Some machine (with a very complex interior in general, see below) is coupled to a thermal reservoir at 
inverse temperature $\beta$ and a work reservoir. The work reservoir can be used to drive the computation in a certain 
direction whereas the thermal reservoir is equipped with a well-defined notion of heat and entropy. The 
task of our machine is to compute. It therefore receives input signals and transforms them to output signals corresponding 
to the result of the computation. Logical reversibility demands that we use two separate tapes for the inputs and 
outputs (henceforth called the input and output tapes, respectively). In fact, if we do not keep the initial inputs but 
simply overwrite them with the output (as the TM from Sec.~\ref{sec TMs} would do), our machine will be in general 
irreversible \footnote{For instance, imagine 
that our machine is designed to add two numbers $a$ and $b$, which are given as their binary equivalent on the input string. 
Clearly, knowing only the result $c$ of the computation, i.e., $c = a + b$, does not allow us to infer the values of $a$ 
and $b$ and hence, our machine would be logically irreversible. }. 

\begin{figure}
 \includegraphics[width=0.45\textwidth,clip=true]{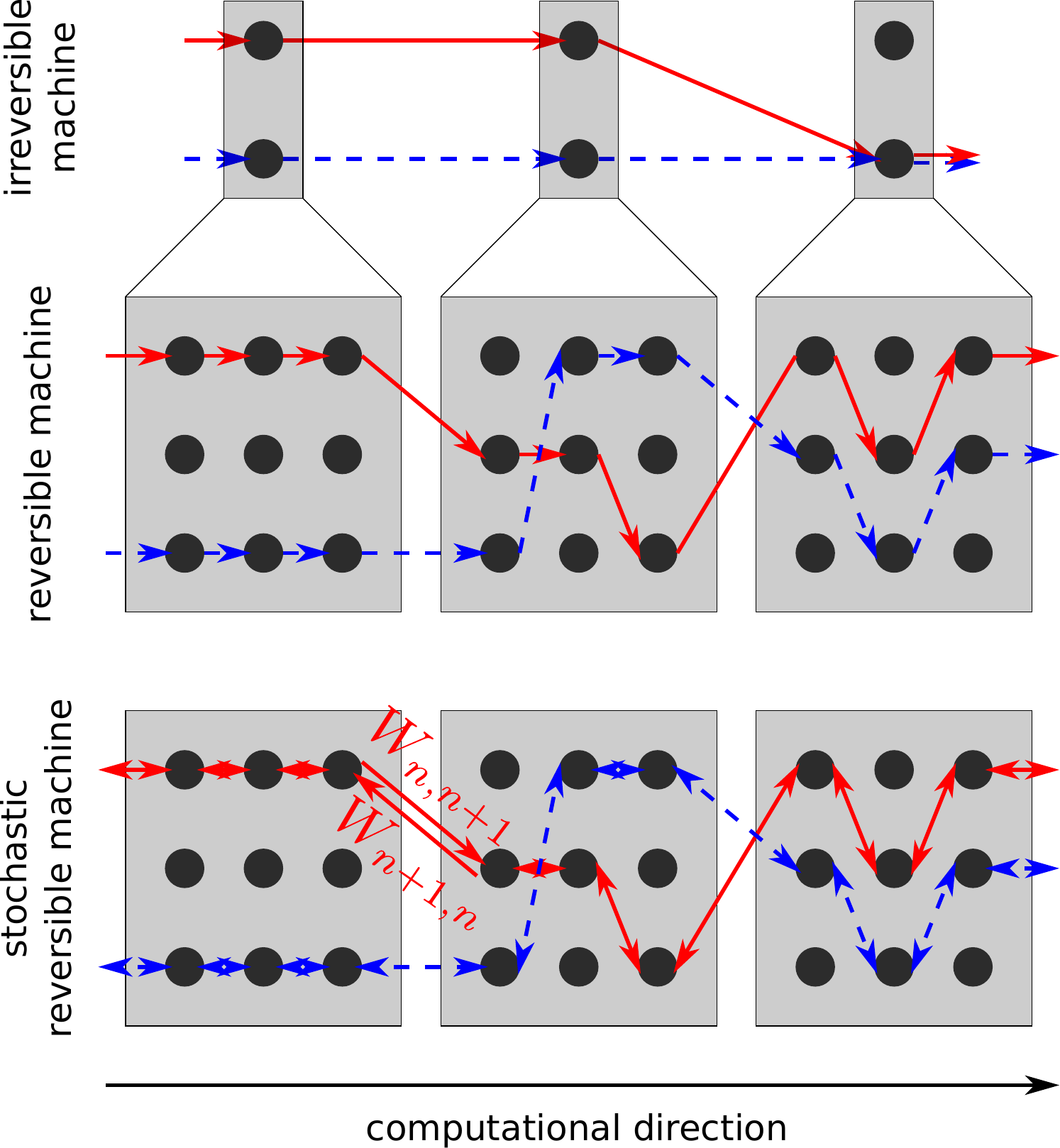}
 \caption{\label{fig comp paths} A computational path is defined by the way the machine proceeds from state to state through a high-dimensional 
 state space (each state of the machine including the tapes is symbolized by a black circle). 
 Here, we sketched two different paths (solid red and dashed blue lines). For standard machines (which compute 
 only from left to right) each path is uniquely determined by the input signal string and the initial machine state. For a 
 logically irreversible machine, however, it might happen that two different paths merge at some point (as shown on top) 
 making it impossible to find the inverse of a state in general 
 (that is to say there is a unique way to go from left to right in the diagram but not from right 
 to left). In contrast, this cannot happen for logically reversible machines (as shown in the middle), but this feature 
 comes at the cost of introducing additional states and tapes making the state space even larger. Finally, whereas 
 the standard logically reversible TM still proceeds deterministically from left to right, a stochastic TM jumps randomly 
 according to some rates [e.g., $W_{n+1,n}$ and $W_{n,n+1}$ as used in Eqs. (\ref{eq ME one step process}) and 
 (\ref{eq help 1})] and hence, it can move both ways (bottom). }
\end{figure}

More specifically, a logically reversible computer is in principle able to unambiguously retrace its \emph{computational 
path} (i.e., the sequence of logical states visited so far) back to the initial state. 
In fact, most TMs as introduced in Sec.~\ref{sec TMs} are logically irreversible, for instance, already due to the fact 
that the machines usually do not remember from which direction they were coming from. 
But even if it would remember this (for instance by writing the direction $d$ 
on an additional tape), the machine might still be logically irreversible. Consider for example that there exists a pair 
of states and symbols $(q_1,s_1)$ and $(q_2,s_2)$ such that $q' = F(q_1,s_1) = F(q_2,s_2)$, $s' = G(q_1,s_1) = G(q_2,s_2)$ 
and $d' = H(q_1,s_1) = H(q_2,s_2)$. Then, given the state $(q',s')$ [or even $(q',s',d')$], the machine has two possible 
predecessor states and it cannot know from which it was coming. Hence, it is logically irreversible. 
This situation is sometimes called the merging of two computational paths, see Fig.~\ref{fig comp paths}. 

However, even a logically reversible computer still proceeds deterministically step by step along its computational path. 
This deterministic behaviour unambigously defines a \emph{computational direction}, see Fig.~\ref{fig comp paths} again 
(we remark that the computational direction does not necessarily coincide with the ``physical'' direction of the movement of 
the tape, which can -- as we have seen -- be shifted either way). In contrast, we also want to look at a stochastic 
machine, which makes random transitions in both directions, i.e., it is also allowed to jump back to a previous computational 
state. In order to assure that we can control the direction of computation \emph{on average}, we demand that we can change 
the potential energy externally, for instance, by using a work source or by adjusting chemical potentials appropriately. 


In the next section we will present the main ideas of how to obtain a logically reversible TM from an irreversible TM as 
discussed in Sec.~\ref{sec TMs} sparing the mathematical details to the appendix. Then, given that we have a logically 
reversible TM, we will show how to associate a ME to it in Sec.~\ref{sec stochastic TM subsec}. 

\subsection{Logically reversible Turing machine}
\label{sec logical reversible TM}

The procedure to make a computation of a TM logically reversible was explained in detail in a famous publication by 
Bennett~\cite{BennettIBM1973}. He explicitly showed -- given an arbitrary TM $T$ as described by 
Eq.~(\ref{eq TM standard map}) -- how to construct a machine $R$ which is computationally equivalent to $T$ but always 
logically reversible. To accomplish this, Bennett introduced a finite number of new machine states and two additional 
tapes, which record the previous computational steps taken. Furthermore, the computation is broken up into three 
\emph{stages} (each stage in general consists of many single computational steps) and in total, the logically reversible 
TM needs approximately four times as many steps as the irreversible one~\cite{BennettIBM1973}. Before we proceed, 
we remark that the construction given by Bennett is, of course, not unique (as he discusses as well) but seems to be very 
convenient. It is also noteworthy that TMs acting on $n$ tapes are not more powerful (in terms of what they can compute) 
than a TM described by Eq.~(\ref{eq TM standard map}) because one can be mapped onto the other~\cite{MinskyBook}. 


In addition to the treatment of Bennett, who considered only a single computation 
$T: \bb s_\text{in} \rightarrow \bb s_\text{out}$, we want to construct a machine which 
continuously processes a stream of incoming input strings of the form 
$(\dots b,\bb s'_\text{in},b,\dots,b,\bb s_\text{in},b,\dots)$. Thus, we imagine an infinite input tape 
with different input strings $\bb s_\text{in}, \bb s'_\text{in}, \dots$ separated by blank symbols to mark the 
beginning and the end of each single input string. The output tape then contains the results of the different computations, 
i.e., it looks like 
$(\dots,b,\bb s'_\text{out} = T(\bb s'_\text{in}),b,\dots,b,\bb s_\text{out} = T(\bb s_\text{in}),b\dots)$. 
In this picture our machine resembles the devices from 
Refs.~\cite{MandalJarzynskiPNAS2012, BaratoSeifertEPL2013, MandalQuanJarzynskiPRL2013, DeffnerJarzynskiPRX2013, BaratoSeifertPRE2014, StrasbergEtAlPRE2014} 
in which also an external tape is manipulated but mainly to extract work and not for computational purposes though. 

 \begin{table*}
  \centering
   \begin{tabular}{llllll}
    stage					&	~~~ input tape ~~~		&	working tape ~~~		&	history tape ~~~		&	output tape ~~~		&	short form, see Eq. (\ref{eq computational cycle})	\\
    \hline 
						&	~~~ input\underline{ }		&	\underline{ }		&	\underline{ }		&	\underline{ }		&	$T_5 T_4 T_3 T_2 T_1(\bb s_\text{in},\bb b,\bb b,\bb b)$	\vspace{0.2cm}	\\
    1) copy input onto working tape 		&					&				&				&				&	\vspace{0.2cm}	\\
						&	~~~ \underline{ }input		&	\underline{ }input	&	\underline{ }		&	\underline{ }		&	$T_5 T_4 T_3 T_2(\bb s_\text{in},\bb s_\text{in},\bb b,\bb b)$	\vspace{0.2cm}	\\
    2) compute					&					&				&				&				&	\vspace{0.2cm}	\\
						&	~~~ \underline{ }input		&	output\underline{ }	&	histor\underline{y}	&	\underline{ }		&	$T_5 T_4 T_3 (\bb s_\text{in},\bb s_\text{out},\bb h,\bb b)$	\vspace{0.2cm}	\\
    3) copy output to output tape		&					&				&				&				&	\vspace{0.2cm}	\\
						&	~~~ \underline{ }input		&	output\underline{ }	&	histor\underline{y}	&	\underline{ }output	&	$T_5 T_4 (\bb s_\text{in},\bb s_\text{out},\bb h,\bb s_\text{out})$	\vspace{0.2cm}	\\
    4) retrace computation			&					&				&				&				&	\vspace{0.2cm}	\\
						&	~~~ \underline{ }input		&	\underline{ }input	&	\underline{ }		&	\underline{ }output	&	$T_5 (\bb s_\text{in},\bb s_\text{in},\bb b,\bb s_\text{out})$	\vspace{0.2cm}	\\
    5) erase working tape			&					&				&				&				&	\vspace{0.2cm}	\\
						&	~~~ \underline{ }input$\star$	&	\underline{ }		&	\underline{ }		&	\underline{ }output	&	$(\bb s_\text{in}\star,\bb b,\bb b,\bb s_\text{out})$	\\
   \end{tabular}
   \caption{\label{table} One computational cycle consists of five stages where each stage can consist of many steps. 
   The first line shows the initial situation of the tapes. 
   After the first stage, the tapes are shown as in the second line, which serves as the initial state for the second stage 
   and so on. The underbar denotes the current position of the machine head on the respective tapes. Note that at each 
   stage the machine works with two tapes only, whereas the other two remain unchanged. Furthermore, in the very last line 
   we used a $\star$ to mark the input the computer has already processed, see appendix~\ref{app logical reversible TM} 
   for details. }
 \end{table*}
 
Our logically reversible TM thus will have in total four tapes and one computational 
cycle proceeds in five stages. The four tapes are called the input, working, history and output tape. Whereas the input 
and output tape are supplied externally (see Fig.~\ref{fig TD comp}), the working and history tape belong 
to the machine itself. By this we especially want to emphasize that we require them to be blank at the end of the 
computation such that they are ready for usage again \footnote{If we do not return the internal tapes of the machine 
back to their initial blank state, we would have to erase them at some point, which dissipates additional heat according 
to Landauer's principle and this needs to be avoided. }. 
More specifically, one computational cycle consists of the following five stages 
(also see Table~\ref{table}): 
\begin{enumerate}
 \item[1)] \emph{Copy input onto working tape:} A new input arrives at the machine on the input tape and the machine copies this 
 input onto its working tape leaving it there in standard format at the end of the first stage. 
 \item[2)] \emph{Compute:} In this stage the actual computation is performed which finally maps the input to the output. 
 Furthermore, a history tape keeps track of the intermediate steps such that the computer would be able to retrace 
 each step. 
 \item[3)] \emph{Copy output to output tape:} If the computation halts, the output on the working tape is copied to the output 
 tape and the working tape is reset to its position as at the end of stage 2. 
 \item[4)] \emph{Retrace computation:} The computer retraces all its computation such that the output on the working tape becomes 
 the input and the history is blank again. This stage is the inverse of stage 2. 
 \item[5)] \emph{Erase working tape:} We erase the working tape with the help of the input tape such that the working tape is 
 blank again. Note that this erasure step is logically reversible because we have an identical copy of the input 
 on the input tape. Finally, we use an additional symbol ($\star$) to mark that we already performed a computation for 
 the current input and the machine moves on to the next input on the input string. 
\end{enumerate}

Stage 2, 3 and 4 were already treated by Bennett in Ref.~\cite{BennettIBM1973}. In addition, we require stages 1 and 5 
because we want that our machine works continuously and not only once. 
The reader who is curious about the details of each step is refered to appendix~\ref{app logical reversible TM}. 
Otherwise, instead of one big machine $T$ doing a computation in five stages, it might also help to imagine 
five small machines $T_1, \dots, T_5$. Our big machine $T$ is then nothing else than a composition or concatenation of 
these small machines, i.e., $T = T_5\circ T_4\circ T_3\circ T_2\circ T_1 \equiv T_5 T_4 T_3 T_2 T_1$ (similar to the 
composition of different functions). The action of $T$ on the state of the four tapes (in the order of the input, working, 
history and output tape, respectively) 
can be written as 
\begin{equation}\label{eq computational cycle}
 \begin{split}
  \text{1) } & T_5 T_4 T_3 T_2 T_1(\bb s_\text{in},\bb b,\bb b,\bb b) = T_5 T_4 T_3 T_2(\bb s_\text{in},\bb s_\text{in},\bb b,\bb b),	\\
  \text{2) } & T_5 T_4 T_3 T_2 (\bb s_\text{in},\bb s_\text{in},\bb b,\bb b) = T_5 T_4 T_3 (\bb s_\text{in},\bb s_\text{out},\bb h,\bb b),	\\
  \text{3) } & T_5 T_4 T_3 (\bb s_\text{in},\bb s_\text{out},\bb h,\bb b) = T_5 T_4 (\bb s_\text{in},\bb s_\text{out},\bb h,\bb s_\text{out}),	\\
  \text{4) } & T_5 T_4 (\bb s_\text{in},\bb s_\text{out},\bb h,\bb s_\text{out}) = T_5 (\bb s_\text{in},\bb s_\text{in},\bb b,\bb s_\text{out}),	\\
  \text{5) } & T_5 (\bb s_\text{in},\bb s_\text{out},\bb b,\bb s_\text{out}) = (\bb s_\text{in}\star,\bb b,\bb b,\bb s_\text{out}).
 \end{split}
\end{equation}
Eq. (\ref{eq computational cycle}) can be regarded as a short form of Table~\ref{table}. Here, $\bb b$ denotes a 
blank tape and $\bb h$ denotes the history tape at the end of stage 2. Clearly, if all the machines 
$T_1, \dots, T_5$ are logically reversible, then also $T$ is.

\subsection{Stochastic Turing machine}
\label{sec stochastic TM subsec}

In this section we will show how to use a continuous-time Markov process to model any logically reversible TM, which we will 
simply call a stochastic TM. We start, however, by repeating what a ME is and what we need for a consistent thermodynamic 
interpretation. 

A continuous-time Markov process describing the dynamics of a system $X$ 
corresponds to a set of states $\C X$ and an associated probability $p_x$ to be in a state $x\in\C X$, which 
changes according to a Markovian first order differential equation called the ME~\cite{VanKampenBook}: 
\begin{equation}\label{eq ME generic}
 \frac{d}{dt}p_x(t) = \sum_{x'} W_{x,x'} p_{x'}(t).
\end{equation}
Here, the rate matrix $W_{x,x'}$ has real-valued entries and fulfills $\sum_x W_{x,x'} = 0$ for all $x'\in\C X$. 
This guarantees that probability is conserved throughout the evolution, i.e., $\frac{d}{dt}\sum_ xp_x(t) = 0$ for all $t$. 
If we want to equip the ME (\ref{eq ME generic}) with a thermodynamic interpretation~\cite{SchnakenbergRMP1976}, 
we have to associate to each state $x$ 
an energy $E_x\in\mathbb{R}$ and the rate matrix has to additionally fulfill a property called \emph{local detailed balance}, 
which states that $\ln[W_{x,x'}/W_{x',x}] = -\beta(E_x-E_{x'})$ where $\beta$ is the inverse temperature of the environment to 
which the system $X$ has contact. Note that this automatically implies that, if $W_{x,x'} \neq 0$, then also $W_{x',x}\neq 0$, 
i.e., for each transition $x' \rightarrow x$ the reversed one $x\rightarrow x'$ must also exist. 
This framework can also be extended to more general situations involving, e.g., multiple 
environments at different temperatures~\cite{SchnakenbergRMP1976}, but we do not need more than the things just mentioned. 

Now, associating a ME to a logically reversible deterministic TM is in fact very easy. All we have to do is to change the 
(unidirectional) deterministic updating rules from appendix~\ref{app logical reversible TM} into (bidirectional) 
probabilistic transition rules, i.e., we allow for transitions in the computational forward direction as well as 
transitions which just undo the last computational step (backward direction). 

It is worth pointing out that the underlying state space $\C X$ of the ME is in general infinite, but this is not necessarily 
related to the size of the input tape. Remember that -- due to the halting problem -- a computation might go on forever 
even if it only received a finite input. In fact, even if the computation halts, there is no general way to give a reasonable 
estimate of the size of $\C X$ in advance~\cite{MinskyBook}.
However, on the other hand, the structure of the ME is very simple and this is 
related to the fact that we build the machine in a logically reversible way. In fact, each state $x\in\C X$ has only two 
adjacent states, namely its logical predecessor and its logical successor state. Hence, our ME describes a simple one-step 
or birth-and-death process~\cite{VanKampenBook} or equivalently, according to Schnakenberg~\cite{SchnakenbergRMP1976}, 
we could say that the topology of 
the underlying network is trivial. In fact, if there were any branchings or loops in the underlying network, the computation 
would not be logically reversible anymore because then a state could have multiple predecessors or successors. Hence, quite 
generally we can put the final ME into the form 
\begin{equation}
 \begin{split}\label{eq ME one step process}
  \frac{d}{dt}p_n(t)	=&	-(W_{n+1,n} + W_{n-1,n})p_n(t)	\\
			&+	W_{n,n+1}p_{n+1}(t) + W_{n,n-1}p_{n-1}(t).
 \end{split}
\end{equation}
Here, of course, $n\in\mathbb{Z}$ is a multi-index denoting the entire machine and tape configuration. 

Let us discuss the general structure of the ME a little further. First of all, it is important to note that only the 
current squares of the tapes can change stochastically whereas the rest of the tapes, which is not 
coupled to the machine, remains fixed. In fact, although there is an infinite number of possible different states, 
not all states $x\in\C X$ are coupled with each other. Which states are coupled to each other is determined by the rules 
from appendix~\ref{app logical reversible TM} and by the input strings $\bb s_\text{in}$ because they single out a unique 
computational path through the ``labyrinth'' of states in $\C X$. 

In addition, the number of transition rules is always \emph{finite} and fixed as expressed in appendix 
\ref{app logical reversible TM}. This is true independently of the number of computations or the lengths of the input 
strings. Although there seem to be quite a lot of rules, note that they suffice to build a 
\emph{universal logically reversible computer}. Of course, things become much easier if we relax some of the 
requirements. Hence, in a more pictorial language we could say that the hardware (i.e., the set of transitions rules 
with the associated rates) of our machine remains fixed, but the software (i.e., the inputs determining the computational 
path) can change.

Thus, if we focus only on the computation for one input string, i.e., on the stages 2 to 4, the full rate matrix 
$W$ in Eq. (\ref{eq ME generic}) decomposes into blocks for each input string $\bb s$, i.e., it has the form 
\begin{equation}
 W = \left(\begin{array}{cccc}
        \boxed{W(\bb s_1)}	&				&				&	\\
				&	\boxed{W(\bb s_2)}	&				&	\\
				&				&	\boxed{W(\bb s_3)}	&	\\
				&				&				&	\ddots 	\\
       \end{array}\right)
\end{equation}
and no transition between different blocks is allowed. Here, we labeled the different input strings 
$\bb s_1, \bb s_2, \bb s_3, \dots$ in some canonical way and note that each block can be infinitely large if the 
computation does not halt. For each $\bb s_i$ the ME describes a simple one-step 
process and by rearranging the states appropriately we can write each block as a tridiagonal matrix 
\begin{widetext}
 \begin{equation}\label{eq help 1}
   W(\bb s_i) = \left(\begin{array}{ccccc}
                        \ddots	&					&				&					&	 	\\
				&	-W_{n,n-1}-W_{n-2,n-1}		&	W_{n-1,n}		&	0				&	 	\\
				&	W_{n,n-1}			&	-W_{n+1,n}-W_{n-1,n}	&	W_{n,n+1}			&	 	\\
				&	0				&	W_{n+1,n}		&	-W_{n+2,n+1}-W_{n,n+1}		&	 	\\
                        	&					&				&	 				&	\ddots 	\\
                       \end{array}\right)
 \end{equation}
\end{widetext}
where $W_{n+1,n}$ ($W_{n-1,n}$) denotes the forward (backward) rate at step $n$ in the ME (\ref{eq ME one step process}). 
Hence, to conclude, although the state space $\C X$ is extremely large, the rate matrix is also extremely 
\emph{sparse}, i.e., it contains only a small number of non-zero elements (in relation to the total number of elements). 

Finally, we would need to associate a consistent energy landscape to our system and the rate of forward and backward 
transitions would then need to obey local detailed balance, which fixes the temperature 
of the environment. We will discuss this issue in the next section.

\section{Thermodynamics of Brownian computation}
\label{sec thermodynamics Brownian computation}

\subsection{Energy landscapes}
\label{sec energy landscapes}

So far we have shown that a stochastic, logically reversible TM can be modeled by a simple one-step process as given by 
the ME (\ref{eq ME one step process}). To interpret it thermodynamically we still need to associate a consistent 
energy landscape to it, which we could control externally via a work source or, alternatively, a bias in 
chemical potentials as it would be the case for biochemical processes. 

For the sake of simplicity, we will choose below a linear energy landscape along the computational path, 
i.e., the difference in energy between a logical state and its successor state is taken 
to be the constant $\epsilon$ (i.e., for $\epsilon > 0$ the computation proceeds on average in the forward direction 
along a chain of states with decreasing energies). This choice is in agreement with the one usually appearing in the 
literature~\cite{BennettIBM1973, Bennett1982, Feynman1985, NortonFoundPhys2013}. Before we proceed, however, we discuss 
and justify this choice in more detail. 

First of all, in Sec.~\ref{sec effective FPE} we will actually discuss the thermodynamics of our model on a coarse-grained 
level of description. 
That is to say we will be interested in the regime where the computer was running already for quite a long time such that 
the variance $\lr{n^2}-\lr n^2$ of the number of computational steps is large compared to unity where we defined 
$\lr{n^\ell} \equiv \sum_n n^\ell p_n(t)$. 
In this picture, $\epsilon$ might denote just an average slope in 
the energy landscape, i.e., we explicitly allow for spatial irregularities in the energy landscape as long as they are 
not too large. More specifially, if $E_n$ denotes the energy of state $n$ according to the ME (\ref{eq ME one step process}), 
we demand that 
\begin{equation}\label{eq epsilon}
 \epsilon \overset{!}{=} \frac{1}{2N}\sum_{n=-N}^{N-1} (E_n-E_{n+1})
\end{equation}
holds for all $n$ and for $N$ of the order of the variance $\lr{n^2}-\lr n^2$ such that the energy landscape looks 
linear at the coarse-grained level. 

Second, it is worth pointing out that in fact -- except for the spatially allowed irregularities -- no other energy 
landscape seems to be feasible for a general purpose computer. The reason 
for this is twofold: first, we are interested in a steady state regime, i.e., the average dissipation \emph{per step} 
should be independent of the number of computational steps performed so far. This demand rules out quadratic or 
exponential energy landscape. Second, we are also still faced with the halting problem. This implies that we cannot know 
in advance the number of computational steps we need for one computational cycle. Thus, associating any particularly 
shaped energy landscapes like a sine or a hill (as in~\cite{NortonIntJModPhys2014}) is unfeasible because we do 
not know, for instance, how to choose a senseful period for the sine. 

Hence, we conclude: the only feasible energy landscape with which we can ensure to control the speed and direction of 
computation independently of the number of computational steps (which we cannot know in advance) and which is 
translationally invariant on the state space $n\in\mathbb{Z}$ of the ME (\ref{eq ME one step process}) is an on average 
linear landscape.

\subsection{Effective Fokker-Planck equation}
\label{sec effective FPE}

Having agreed on the (on average) linear energy landscape we will choose the transition rates in Eq. 
(\ref{eq ME one step process}) as follows: 
\begin{equation}
 W_{n,n+1} = \Gamma e^{-\beta\epsilon/2}, ~~~ W_{n+1,n} = \Gamma e^{\beta\epsilon/2}.
\end{equation}
Here, $\Gamma$ is some rate setting the overall time-scale of our problem and we see that the rates fulfill local 
detailed balance, i.e., $\ln[W_{n,n+1}/W_{n+1,n}] = -\beta (E_n-E_{n+1}) = -\beta\epsilon$ where $\beta\epsilon>0$ 
favors a computation in the forward direction. 

In the limit where the mean $\langle n\rangle$ and variance $\lr{n^2}-\lr n^2$ are large compared to one, we can 
approximate derivatives by 
\begin{equation}
 \begin{split}
  \frac{\partial}{\partial n}p_n(t)	&\approx		\frac{p_{n+1}(t)-p_{n-1}(t)}{2},	\\
  \frac{\partial^2}{\partial n^2}p_n(t)	&\approx		p_{n+1}(t)-2p_n(t)+p_{n-1}(t).
 \end{split}
\end{equation}
Then, the Fokker-Planck equation (FPE) corresponding to the ME (\ref{eq ME one step process}) reads 
\begin{equation}\label{eq FPE}
 \frac{1}{\Gamma}\frac{\partial}{\partial t}p_n(t) = \frac{\partial}{\partial n}\left[-2\sinh\frac{\beta\epsilon}{2} + \cosh\frac{\beta\epsilon}{2}\frac{\partial}{\partial n}\right] p_n(t).
\end{equation}
This FPE describes the movement of an overdamped Brownian particle in a constant force field and with a constant diffusion 
coefficient with $n\in\mathbb{R}$ denoting the position of the particle. It even admits an explicit solution: 
assuming that the machine has started initially at some fixed state $n=0$, i.e., $p_n(t=0) = \delta(n)$, we obtain 
\begin{equation}\label{eq FPE solution}
 \begin{split}
  p_n(t)	=&~	\frac{1}{\sqrt{4\pi\Gamma\cosh(\beta\epsilon/2)t}}	\\
		&\times	\exp\left\{-\frac{[n-2\Gamma\sinh(\beta\epsilon/2)t]^2}{4\Gamma\cosh(\beta\epsilon/2)t}\right\}.
 \end{split}
\end{equation}

\subsection{Thermodynamic discussion}

Discussing the thermodynamic behaviour of Eq. (\ref{eq FPE}) can be done using standard methods, see e.g. Ref. 
\cite{SekimotoBook}. We first of all compute the mean number of computational steps, which is 
\begin{equation}\label{eq average n}
 \lr n(t) = 2 \Gamma t \sinh\frac{\beta\epsilon}{2}
\end{equation}
and hence, the \emph{speed} of computation becomes 
\begin{equation}
 v \equiv \frac{d}{dt}\lr n(t) = 2 \Gamma \sinh\frac{\beta\epsilon}{2},
\end{equation}
which -- as expected -- can be controlled by $\epsilon$. Especially, we see that 
we have $v > 0$ for $\epsilon > 0$ and vice versa. 

The variance of the distribution is 
\begin{equation}
 \lr{n^2}(t) - \lr n^2(t) = 2 \Gamma t \cosh\frac{\beta\epsilon}{2}
\end{equation}
and grows linearly with time. Based on this we might ask the question when does the computation become approximately 
deterministic, i.e., when does the mean dominate the standard deviation? Calculating their ratio yields 
\begin{equation}
 \frac{\lr n(t)}{\sqrt{\lr{n^2}(t) - \lr n^2(t)}} = \sqrt{2\Gamma t \tanh\frac{\beta\epsilon}{2}\sinh\frac{\beta\epsilon}{2}}.
\end{equation}
If we want this quantity to be much larger than one, we obtain a condition for the minimum amount of time we have to wait 
until our machine computes almost in a deterministic fashion: 
\begin{equation}
 t \gg \left(2\Gamma\tanh\frac{\beta\epsilon}{2}\sinh\frac{\beta\epsilon}{2}\right)^{-1} \approx \frac{2}{\Gamma\beta^2\epsilon^2}.
\end{equation}
Here, we performed an expansion in $\beta\epsilon\ll1$ at the end. Thus, the closer we get to the reversible limit, i.e., 
the smaller $\beta\epsilon$ (see below), the longer we have to wait until the computer starts to work reliably. 

Furthermore, we can explicitly calculate the Shannon entropy of our distribution, which is 
\begin{equation}
 H(t) \equiv -\int dn p_n(t)\ln p_n(t) = \frac{1}{2}\ln\left(4\pi e \Gamma t \cosh\frac{\beta\epsilon}{2}\right).
\end{equation}
Using Eq. (\ref{eq average n}) we can also write the Shannon entropy as 
\begin{equation}
 H(t) = \frac{1}{2}\ln\left[2\pi e \coth\left(\frac{\beta\epsilon}{2}\right) \lr n(t)\right],
\end{equation}
i.e., the Shannon entropy scales with the average number $\lr n$ of computational steps as $S(t) \sim \ln\lr n$. 

The rate at which entropy is produced is given by the change in Shannon entropy plus $\beta$ times the heat flow 
dissipated into the environment~\cite{SekimotoBook}. Since the latter is simply $\epsilon v$, we can write for the 
entropy production rate 
\begin{equation}
 \begin{split}
  \dot S_\bb{i}(t)	&=	\frac{d}{dt} H(t) + \beta\epsilon v	\\
			&=	\frac{1}{2t} + 2\Gamma\beta\epsilon \sinh\frac{\beta\epsilon}{2} \ge 0.
 \end{split}
\end{equation}
We now note that for $\epsilon \rightarrow 0$ the last term vanishes quadratically, i.e., the heat dissipated can be 
made arbitrarily small in this limit. The first term, however, is independent of $\epsilon$ but vanishes for 
$t\rightarrow\infty$. Hence, we have 
\begin{equation}
 \lim_{t\rightarrow\infty} \lim_{\epsilon\rightarrow0} \dot S_\bb{i}(t) = 0.
\end{equation}
Thus, a Brownian computer can work thermodynamically reversibly (i.e., with zero entropy production rate) 
in the steady state regime if the bias $\epsilon$ is small enough. This would confirm the conclusions from Ref. 
\cite{BennettIBM1973, Bennett1982, Feynman1985}. 

However, we can also confirm Norton's perspective on the matter~\cite{NortonFoundPhys2013, NortonIntJModPhys2014}. 
Starting initially at $t=0$ we see that the total amount of entropy produced up to time $t$ is 
\begin{equation}
 \Delta_\bb{i} S(t) = \int_0^t dt' \dot S_\bb{i}(t') = H(t) + \beta \epsilon \lr n(t) \ge 0.
\end{equation}
Thus, even for $\epsilon = 0$ the Shannon entropy still grows logarithmically with the number of computational steps 
because the probability distribution of the machine spreads over the available phase space similarly to the 
free expansion of a one-molecule gas, which is a thermodynamically irreversible process. 
If we think in terms of thermodynamic cycles instead of a thermodynamic machine, which works in the stationary regime, 
we would have to dissipate an amount of entropy proportional to $\ln\lr n$ to reset the Brownian computer to its 
initial zero entropy state. However, again, compared to the number of computational steps $\lr n$ taken, the ratio 
$\ln(\lr n)/\lr n$ becomes arbitrarily small for a large number of steps, i.e., for a long computation. 

Finally, we remark that a full treatment in terms of the ME (\ref{eq ME one step process}) instead of the FPE 
(\ref{eq FPE}) would provide very similar results. In fact, the first two cumulants (mean and variance) can be shown 
to coincide. 

\section{Summary and outlook}
\label{sec discussion}

Let us summarize our main findings and discuss possible interesting open questions based on our findings. 

We have started with an arbitrary TM as a model for a general purpose computer, which is, however, in general logically 
irreversible and free from any thermodynamic interpretation. We then discussed how a computation can be regarded as a 
thermodynamic process and we decided to investigate the thermodynamics of a logically reversible stochastic computer, 
which simulates the TM considered at the beginning. We explained in detail how to obtain a logically reversible computer 
out of an irreversible one and then used bidirectional probabilistic transition rules instead of unidirectional 
deterministic rules to model the stochastic or random motion of our computer. 

Although the problem seems to be very complex, we have seen that the resulting ME has the very simple structure of a 
one-step process, which describes how the computer follows a well-defined path in a large state space and eventually 
maps the input signals to output signals (the results of the computation). 
We have argued that the only feasible energy landscape for such a computer is an approximately 
linear one, which not only simplified the thermodynamic discussion, but also allowed us to solve a corresponding 
FPE describing the drift and diffusion of the computer exactly. 

We have then seen that our stochastic computer can work thermodynamically reversibly, i.e., in a dissipation-free fashion, 
in a steady 
state regime and in this respect Feynman was indeed right with his initially quoted statement. However, just because the 
entropy production rate can become arbitrary small, this does not imply that the overall integrated entropy production is 
zero. Especially, if we think in terms of a computational cycle, in which we want to reset the computer to its initial state 
at the end, there is an unavoidable cost due to the increasing Shannon entropy of the probability distribution during the 
computation. In fact, this additional cost is \emph{not} independent of the number of computational steps but scales 
logarithmically with it and it seems that this effect has been only recognized by Norton so far~\cite{NortonFoundPhys2013}. Here, 
we have verified this result in a conceptually clean and general framework. 

Furthermore, it is worth emphasizing that our computer works \emph{error-free} at a \emph{finite} entropy production. 
In fact, by construction our model does only allow for temporary errors in the computation (due to the fact that our 
stochastic machine can randomly hop back to its previous state), but in the long run each temporary error is corrected 
by the next step in the computational forward direction and there are no other sources of errors allowed. Including 
errors (for instance, random bit flips or -- to avoid the halting problem and an infinitely long computation -- one could 
decide to terminate the computation after a fixed number $N_\text{max}$ of steps) 
in our scheme and investigating the thermodynamic cost to correct them might be an interesting project for the 
future. 

Another interesting question is whether we can sensefully assign a notion of \emph{efficiency} to our computer. From a 
purely physical point of view the machine we have considered is actually senseless because it describes only a simple 
(and never-ending) relaxation process. However, the machine is indeed ``working'', i.e., doing something ``useful'' for 
us, because it tells us the answer to many questions. But how can we quantify the usefulness of our 
machine? Having a rigorous notion of a thermodynamic efficiency for a computer would allow us to study question of, e.g., 
efficiency at maximum power, which is an important question for the design of realistic machines, see, 
e.g.,~\cite{CurzonAhlbornAJP1975, VandenBroeckPRL2005, SchmiedlSeifertEPL2008, EspositoLindenbergVandenBroeckPRL2009, EspositoLindenbergVandenBroeckEPL2009}. 
In this context, one can also ask the question whether a logically reversible computer is really desirable or whether 
a logically irreversible computer might indeed be able to work at a fundamentally better efficiency.
At the end, it seems that biochemical processes in our body 
work very efficiently, but not necessarily in a logically reversible way. 

Finally, let us say a few words about the relation between the present work and the devices investigated in 
Refs.~\cite{MandalJarzynskiPNAS2012, BaratoSeifertEPL2013, MandalQuanJarzynskiPRL2013, DeffnerJarzynskiPRX2013, BaratoSeifertPRE2014, StrasbergEtAlPRE2014} 
(also see Feynman for a simple example of such a device who called them information-driven engines~\cite{Feynman1985}). 
Indeed, as in our case, these information-driven engines are coupled to an external tape or ``information reservoir''. 
This additional reservoir can then be used to extract work from a single heat bath while simultaneously writing information 
on the tape (i.e., increasing its Shannon entropy). This picture, however, does not carry over to our situation. 
In fact, the Shannon entropies of the incoming and outgoing tapes are \emph{equal} 
because the input tape gets mapped to itself and the output tape is uniquely determined by the input. 
For a logically irreversible computer this does not need to be true as it can be already seen from the devices 
in 
Refs.~\cite{MandalJarzynskiPNAS2012, BaratoSeifertEPL2013, MandalQuanJarzynskiPRL2013, DeffnerJarzynskiPRX2013, BaratoSeifertPRE2014, StrasbergEtAlPRE2014} 
where it was also shown that they can be used as an information eraser.

\section*{Acknowledgments}

Financial support by the DFG (SCHA 1646/3-1, SFB 910, and GRK 1558) is gratefully acknowledged.


\appendix

\section{Logical reversible Turing machine}
\label{app logical reversible TM}

We here provide the detailed rules for the machine behaviour at each stage of the computation. These rules are, of course, 
not unique. However, because we are not primarily interested in the speed or efficiency of our machine, but only in 
\emph{what} it can do, possibly different implementations are unimportant for the present context. Furthermore, note that 
at each stage the machine is only manipulating two tapes whereas the other two tapes remain fixed (see Table~\ref{table}). 
We will therefore use the notation $[q^{(i)},s_m,t_n]$ where $q^{(i)}$ denotes the internal machine state at stage $i$, 
$s_m$ the symbol $s$ printed on square $m$ of the first tape of interest and $t_n$ the symbol $t$ printed on square $n$ of 
the second tape of interest. What are the tapes of interest will become clear in the treatment of each stage. Note that 
the  notation is different from the one used by Bennett~\cite{BennettIBM1973}.

\subsection{Stage 1) copy input onto working tape}

We want to copy the input on the input tape (first tape of interest) to the working tape (second tape of interest). The 
input is given in the form $\bb s_\text{in} = (s_1,\dots,s_M)$ and we assume that the machine scans initially the symbol 
at the far right (i.e., $s_M$) (see also Table~\ref{table}). Furthermore, the working tape is by construction initially 
completely blank (that this is so can only be seen after the completion of all five stages, of course). The copy stage 
then proceeds as follows: 
\begin{equation}\label{eq stage 1 explicit}
 \begin{split}
  [q_0^{(1)}, (s_M)_m, b_n]	\rightarrow&~	[q_1^{(1)}, (s_M)_m, (s_M)_n]	\\
				\rightarrow&~	[q_0^{(1)}, (s_{M-1})_{m-1}, b_{n-1}]	\\
				\rightarrow&~	[q_1^{(1)}, (s_{M-1})_{m-1}, (s_{M-1})_{n-1}]	\\
				&~		\vdots 	\\
				\rightarrow&~	[q_1^{(1)}, (s_1)_{m-(M-1)}, (s_1)_{n-(M-1)}]	\\
				\rightarrow&~	[q_0^{(1)}, b_{m-M}, b_{n-M}]	\\
				\rightarrow&~	[\text{R}^{(2)},b_{m-M}, b_{n-M}].
 \end{split}
\end{equation}
Hence, we see that during the copy operation the machine changes between the two states $q_0^{(1)}$ and $q_1^{(1)}$ 
where the first is responsible for copying the symbol on the input tape to the working tape and the second is responsible 
for a shift of both tapes. This procedure goes on until it hits the first blank symbol to the left of 
the input string. The machine then changes to the ``ready'' state R$^{(2)}$, which is the special initial state for the 
second stage. Note that -- due to the fact that the copying procedure is unidirectional, i.e., the machine moves the tape 
always in the same direction -- each map has a clear logical inverse. Furthermore, the positions $m$ and $n$ of the 
squares are in general arbitrary and can be chosen initially at will.

\subsection{Stage 2) compute}

This is the central stage of our computational cycle. If we would not bother about logical reversibility, 
this would be the only stage to execute. We thus have to explicitly think about how to make the map 
(\ref{eq TM standard map}) logically reversible. The idea is the following~\cite{BennettIBM1973}: first, because 
each standard (i.e., irreversible) TM is defined by its $N_{\C Q}(N_{\C S}+1)$ many quintuples, we introduce a set 
$\tilde Q$ with $\#\tilde Q = N_{\C Q}(N_{\C S}+1)$ of additional machine states $\tilde q\in\tilde Q$. Then, to each map 
$(q,s) \rightarrow (q',s',d')$ we can associate a special state $\tilde q_{qs}$, which \emph{uniquely} labels each 
quintuple. Second, we will make use of an additional tape called the history tape, which remembers the past 
$\tilde q_{qs}$ such that the machine is able to uniquely retrace its computational path. 

Stage 2 is thus a bit more complicated. The corresponding maps are 
\begin{equation}\label{eq stage 2 explicit}
 \begin{split}
  [\text{R}^{(2)}, b_{n-M}, b_m]&~~~\rightarrow	[\tilde q^{(2)}_{R^{(2)}b}, b_{n-M}, b_{m+1}]	\\
				&~~~\rightarrow	[q_1, (s_1)_{n-M+1}, (\tilde q_{\text{R}^{(2)}b})_{m+1}]	\\
				& ~~~ ~~~ ~~	\vdots 	\\
  \text{step } \ell		&\left\{\begin{array}{ll}
                                  	 \rightarrow	&	[q_\ell^{(2)},(s_\ell)_{n'},(\tilde q_{q_{\ell-1}^{(2)}s_{\ell-1}})_{m+\ell}] \\
                                  	 \rightarrow	&	[\tilde q^{(2)}_{q_{\ell}^{(2)}s_{\ell}},(s'_\ell)_{n'},b_{m+\ell+1}] \\
                                  	 \rightarrow	&	[q'^{(2)}_{\ell},(s_{\ell+1})_{n'+d'},(\tilde q_{q_{\ell}^{(2)}s_{\ell}})_{m+\ell+1}] \\
					\end{array}\right.	\\
                                & ~~~ ~~~ ~~	\vdots 	\\
  \text{step } \nu		&~~~\rightarrow	[\tilde q^{(2)}_{q_{\nu}^{(2)}s_{\nu}},(s'_{\nu})_{n''-1},b_{m+\nu+1}]	\\
				&~~~\rightarrow	[\text{H}^{(2)},b_{n''},(\tilde q_{q_\nu^{(2)}s_\nu})_{m+\nu+1}].
 \end{split}
\end{equation}
The most important step to understand is the one in the middle, which corresponds to the $\ell$'th computational step 
of the ordinary irreversible TM defined by (\ref{eq TM standard map}). Initially, the machine is in state 
$q_\ell^{(2)}$ and scans the symbol $s_\ell$ on the square $n'$ of the working tape. The history tape contains 
the state $\tilde q^{(2)}_{q_{\ell-1}^{(2)}s_{\ell-1}}$, which uniquely labels the \emph{previous} computational step. 
Then, the machine changes its state to $\tilde q^{(2)}_{q_{\ell}^{(2)}s_{\ell}}$, writes the symbol $s'_\ell$ on the 
square according to the function $s'_\ell = G(q^{(2)}_\ell,s_\ell)$ and shifts the history tape one square to the left, 
which contains a blank symbol. Finally, we write $\tilde q_{q_{\ell}^{(2)}s_{\ell}}$ to the history tape and 
change the machine state to $q'^{(2)}_\ell = F(q^{(2)}_\ell,s_\ell)$. Furthermore, we shift the working tape one 
square according to $d' = H(q^{(2)}_\ell,s_\ell)$ such that the machine now scans the new symbol $s_{\ell+1}$. Then, 
the whole procedure can start again where -- in order that we are able to apply map (\ref{eq TM standard map}) -- 
we identify $q'^{(2)}_\ell \equiv q^{(2)}_{\ell+1}$ because the final 
state of the machine at the end of step $\ell$ is the initial state for step $\ell+1$. 

The first and last two lines of Eq. (\ref{eq stage 2 explicit}) then simply describe the initial and final steps of the 
computation. Initially, the machines starts in R$^{(2)}$ and then shifts the working tape to the left such that it reads 
the first symbol $s_1$ and starts in the state $q_1^{(2)}$. Finally, if the machine halts, it reaches the state H and stops 
at the first blank to the right of the output of the computation.

\subsection{Stage 3) copy output to output tape}

We now want to copy the output $\bb s_\text{out} = (s_1,\dots,s_{M'})$ of the computation from stage 2 from the working 
tape (first tape of interest) to the output tape (second tape of interest). This goes as follows 
\begin{equation}
 \begin{split}
  [\text{H}^{(2)}, b_{n''}, b_m]	\rightarrow&~	[q_0^{(3)}, b_{n''}, b_m]	\\
				\rightarrow&~	[q_1^{(3)}, (s_{M'})_{n''-1}, b_{m-1}]	\\
				\rightarrow&~	[q_0^{(3)}, (s_{M'})_{n''-1}, (s_{M'})_{m-1}]	\\
				&~		\vdots 	\\
				\rightarrow&~	[q_0^{(3)}, (s_1)_{n''-M'}, (s_1)_{m-M'}]	\\
				\rightarrow&~	[q_1^{(3)}, b_{n''-M'-1}, b_{m-M'-1}]
 \end{split}
\end{equation}
and is very similar to (\ref{eq stage 2 explicit}). Finally, however, to prepare the machine for the next stage, we want 
that it scans the output on the working tape at the very right again (at the moment it scans the blank on the working tape 
to the left of the output). To accomplish this we use two more machine states: 
\begin{equation}
 \begin{split}
  & [q_1^{(3)}, b_{n''-M'-1}, b_{m-M'-1}]	\\
  & \rightarrow	[q_2^{(3)}, b_{n''-M'-1}, b_{m-M'-1}]	\\
  & \rightarrow	[q_3^{(3)}, (s_1)_{n''-M'}, b_{m-M'-2}]	\\
  & \rightarrow	[q_3^{(3)}, (s_2)_{n''-M'+1}, b_{m-M'-3}]	\\
  & ~~~~~ \vdots 	\\
  & \rightarrow	[q_3^{(3)}, (s_{M'})_{n''-1}, b_{m-2M'}]	\\
  & \rightarrow	[q_3^{(3)}, b_{n''}, b_{m-2M'-1}].
 \end{split}
\end{equation}
Here, $q_2^{(3)}$ is an intermediate state and its sole purpose is to indicate that the copying procedure is over and the 
machine starts now only to shift the working tape without changing it. The state $q_3^{(3)}$ then actually accomplishes 
this task by shifting the working tape one step to the left while simultaneously shifting the output tape one step 
to the right until it reaches the first blank symbol to the right of the output on the working tape, which will indicate 
the start of stage 4.

\subsection{Stage 4) retrace computation}

In this stage we will basically apply the inverse of stage 2 such that at the end the working tape contains the input again 
and the history tape is returned to its initial blank state. This goes as follows: 
\begin{equation}
 \begin{split}
  & [q_3^{(3)}, b_{n''}, (\tilde q_{q_\nu^{(2)}s_\nu})_{m+\nu}]	\\
  & \rightarrow	[\text{H}^{(4)}, b_{n''}, (\tilde q_{q_\nu^{(2)}s_\nu})_{m+\nu}]	\\
  & \rightarrow	[\tilde q^{(4)}_{q_\nu^{(2)}s_\nu}, (s'_\nu)_{n''-1},b_{m+\nu}]	\\
  & \rightarrow	[q_\nu^{(4)}, (s'_{\nu-1})_{n''-1},(\tilde q_{q_{\nu-1}^{(2)}s_{\nu-1}})_{m+\nu-1}]	\\
  & ~~~~~ \vdots 	\\
  & \rightarrow	[q'^{(4)}_\ell, (s_{\ell+1})_{n'+d'}, (\tilde q_{q_\ell^{(2)}s_\ell})_{m+\ell+1}]	\\
  & \rightarrow	[\tilde q^{(4)}_{q_\ell^{(2)}s_\ell}, (s'_\ell)_{n'}, b_{m+\ell+1}]	\\
  & \rightarrow	[q_\ell^{(4)}, (s_\ell)_{n'}, (\tilde q_{q_{\ell-1}^{(2)}s_{\ell-1}})_{m+\ell}]	\\
  & ~~~~~ \vdots 	\\
  & \rightarrow	[q_1^{(4)}, (s_1)_{n-M+1}, (\tilde q_{\text{R}^{(2)}b})_{m+1}]	\\
  & \rightarrow	[\tilde q^{(4)}_{\text{R}^{(2)}b}, b_{n-M}, b_{m+1}]	\\
  & \rightarrow	[\text{R}^{(4)}, b_{n-M}, b_m].
 \end{split}
\end{equation}
Note that we are using the superscript $(4)$ on the \emph{internal} machine states to explicitly distinguish them from the 
states of stage 2 indicating that we are truly in a different stage here.

\subsection{Stage 5) erase working tape}

The last step consists in erasing the content on the working tape such that it is blank again and ready for the next 
computation. Note that the ``erasure'' of the working tape does not actually erase any information because the working 
tape contains the same input $\bb s_\text{in} = (s_1,\dots,s_M)$ as the input tape. As in stage 1 we choose the first 
tape of interest to be the input tape and the second tape of interest is the working tape. Then, we actually only have 
to apply the inverse of stage 1, i.e., 
\begin{equation}
 \begin{split}
  [\text{R}^{(4)}, b_{m-M}, b_{n-M}]	\rightarrow&~	[q_0^{(5)},b_{m-M}, b_{n-M}]	\\
					\rightarrow&~	[q_1^{(5)},(s_1)_{m-(M-1)}, (s_1)_{n-(M-1)}]	\\
					\rightarrow&~	[q_0^{(5)},(s_1)_{m-(M-1)}, b_{n-(M-1)}]	\\
					&~		\vdots 	\\
					\rightarrow&~	[q_1^{(5)},(s_M)_{m}, (s_M)_{n}]	\\
					\rightarrow&~	[q_0^{(5)},(s_M)_{m}, b_{n}]	\\
					\rightarrow&~	[q_1^{(5)},b_{m+1}, b_{n+1}].
 \end{split}
\end{equation}
If we would postulate the final transition rule $[q_1^{(5)},b_{m+1}, b_{n+1}] \rightarrow [q_0^{(1)}, (s_M)_m, b_n]$ we 
would be exactly back at the initial state of stage 1 and the only change would be that we have printed the result of 
the computation on the output tape. However, if this were true, we would be doomed to repeat the same computation again, 
while in fact we want to compute with the \emph{next} input $\bb s'_\text{in}$ on the input string. To achieve this 
we add the following rules: 
\begin{equation}
 \begin{split}
  [q_1^{(5)},b_{m+1}, b_{n+1}]	\rightarrow&~	[q_2^{(5)}, \star_{m+1}, b_{n+1}]	\\
				\rightarrow&~	[q_2^{(5)}, (s_M)_{m}, b_{n+2}]	\\
				\rightarrow&~	[q_2^{(5)}, (s_{M-1})_{m-1}, b_{n+3}]	\\
				&~		\vdots 	\\
				\rightarrow&~	[q_2^{(5)}, (s_{1})_{m-M}, b_{n+M+1}]	\\
				\rightarrow&~	[q_2^{(5)}, b_{m-M-1}, b_{n+M+2}]	\\
				\rightarrow&~	[q_3^{(5)}, b_{m-M-1}, b_{n+M+2}]	\\
				&~		\vdots	\\
				\rightarrow&~	[q_3^{(5)}, (s'_{M'})_{m'}, b_{\hat n}]	\\
				\rightarrow&~	[q_0^{(1)}, (s'_{M'})_{m'}, b_{\hat n}].
 \end{split}
\end{equation}
Here, we first of all marked the input $\bb s_\text{in}$ with a $\star$ to indicate that we have done already a computation 
for that input. We then used an additional state $q_2^{(5)}$, which simply traverses the input string $\bb s_\text{in}$ 
until it hits a blank symbol. The machine then changes to the state $q_3^{(5)}$ and goes further to the left until it 
hits the next \emph{non-blank} symbol on the input string. This symbol then indicates the beginning of the next input 
$\bb s'_\text{in}$ such that -- starting from stage 1 again -- we can readily execute the next computational cycle.

\subsection{Summary}

Suppose that the irreversible TM from Sec.~\ref{sec TMs} has $N_{\C Q}$ many internal states, 
$N_{\C S} + 1$ many different symbols (including the blank) on the tape and hence, it has 
$N_{\tilde Q} \equiv N_{\C Q}(N_{\C S} + 1)$ many 
quintuples. Furthermore, suppose it was given an input of length $M$ and produced an output of length $M'$ after 
$\nu$ computational steps in total. 

Then, our reversible machine has $2 + (N_{\C Q} + N_{\tilde Q}) + 4 + (N_{\C Q} + N_{\tilde Q}) + 4$ states 
from the first, second, ..., fifth stage of the computation, i.e., in total $2(N_{\C Q} + N_{\tilde Q}) + 10$ 
states. Furthermore, it needs $2M + 2\nu + (2M'+1) + 2\nu + (2M+2+x) = 4\nu + 4M + 2M' + 3 + x$ many 
computational steps. Here, the $x$ denotes the number of unknown blank symbols separating the current input from the next 
input on the input tape (see stage 5). Note that we need $x\ge 2$ such that there is enough space for the symbol $\star$ 
and to guarantee that all input strings (including potentially the symbol $\star$) are separated by at least one blank 
symbol from eachother.

\end{document}